# Properties of charge-exchange giant spin-dipole resonances in medium-heavy closed-shell parent nuclei: a semimicroscopic description.

V.I. Bondarenko[1], M.H. Urin[2]

[1] Shubnikov Institute of Crystallography, Federal Research Center "Crystallography and Photonics," Russian Academy of Sciences, Moscow, 119333, Russia;

[2] National Research Nuclear University "MEPhI", Moscow, 115409, Russia

The semimicroscopic particle-hole dispersive optical model is adapted for a description of main properties of charge-exchange giant spin-multipole resonances in medium-heavy closed-shell parent nuclei. The adapted model is implemented to evaluating the strength functions, transition densities, and branching ratios of direct one-nucleon decay for charge-exchange giant spin-dipole resonances in the $^{48}$Ca, $^{90}$Zr, $^{132}$Sn, and $^{208}$Pb parent nuclei. Some of calculation results are compared with available experimental data.

## I. Introduction

Being related to high-energy collective nuclear excitations of particle-hole type, giant resonances (GRs) are the object of extended experimental and theoretical studies (see, e. g., monograph [1]). Each from a great variety of GRs is characterized by the following main energy-averaged quantities: the strength functions, transition densities, and probabilities of direct one-nucleon decay. To evaluate the mentioned GR characteristics, it is necessary to employ a nuclear model, allowing take together into account the main relaxation modes of particle-hole-type states associated with GRs. These modes are: Landau damping, coupling mentioned states to the single-particle (s-p) continuum and to many-quasiparticle configurations (the spreading effect).

In the last decade, the semimicroscopic particle-hole (p-h) dispersive optical model (PHDOM) has been proposed [2] and then adapted to describe the above-mentioned characteristics of various GRs in medium-heavy closed-shell nuclei (Refs. [3, 4] and references therein). PHDOM is a microscopically-based extension of standard and non-standard continuum-RPA (cRPA) versions to taking the spreading effect into account. Within the model, Landau damping and coupling (p-h)-type states to the s-p continuum are described microscopically in terms of a realistic partially self-consistent mean field and Landau-Migdal p-h interaction, while the spreading effect is treated phenomenologically via the energy-averaged p-h self-energy term. The real part of this term is determined by the imaginary part via a proper dispersive relationship. Within PHDOM, the GR characteristics are described in a wide excitation-energy interval, including distant "tails" of considered GRs.

In this paper, we adapt PHDOM to a description of main properties of charge-exchange (isovector) giant spin-multipole resonances (IVGSMPR$^{(\mp)}$s) in medium-heavy closed-shell parent nuclei. As a stage in the PHDOM-based systematic study of the mentioned GRs, we employ the adapted model to evaluate the main characteristics of the charge-exchange giant spin-dipole resonances (IVGSDR$^{(\mp)}$) and their overtones (IVGSDR$^{(\mp)}$2) in the $^{48}$Ca, $^{90}$Zr, $^{132}$Sn, and $^{208}$Pb parent nuclei. This work is a continuation of the recent study of Ref. [4], where the PHDOM-based description of Gamow-Teller and charge-exchange giant spin-monopole resonances in the mentioned parent nuclei is given. Moreover, all the model parameters specified in Ref. [4] are used in the present work. The employed model is a deep modification of the rather old study of Ref. [5], where an intuitive method of cRPA extension to taking the spreading effect into account has been used. Also, we mention a possibility to describe, within the properly extended PHDOM version, the effect of tensor correlations on formation of IVGSMPR$^{(\mp)}$s. The first step in this direction (in applying to Gamow-Teller resonance and its overtone in $^{208}$Bi) has been recently done [6]. Here, we point out special interest to studying the IVGSDR$^{(\mp)}$ strength functions, which are related to the neutron skin in the respective parent nucleus via the non-energy-weighted sum rule. In this work, we also use (in spirit of the study of Ref. [7]) the method, allowing specify, within the model, evaluation of the partial branching ratios of GR direct one-nucleon decay, to make a comparison of these ratios with respective experimental data more adequate.

Concluding the Introduction, we mention RPA-based self-consistent approaches, employing versions of Skyrme-type forces. In particular, these microscopic approaches have been implemented to a description of strength functions of IVGSDR$^{(-)}$ in a few closed-shell parent nuclei [8, 9]. In this description, the spreading effect is simulated by an artificial "smearing procedure". Besides, other characteristics of this GR (transition densities, probabilities of direct one-nucleon decay), properties of the overtone-GR were not considered within these approaches. Being not fully self-consistent semimicroscopic model, PHDOM demonstrates unique abilities in a realistic description of main properties of various GRs in medium-heavy closed-shell nuclei (examples are given in Refs. [3, 4]). We also note, that PHDOM is not directly related to the well-known single-quasiparticle dispersive optical model (see, e. g., Refs. [10, 11] and references therein]). Similarity in formulation of these models consists in describing the spreading effect (phenomenologically and in average over the excitation energy) via, respectively, the p-h and s-p self-energy terms, which are independent of one another. Being averaged over a relatively small energy-interval, which includes many chaotic states, evaluated within these models characteristics of simple modes of high-energy nuclear excitations are independent of this interval. It is noteworthy, that energy density of chaotic (many-quasiparticle) states is described by statistical models.

The paper is organized as follows. In Sect. II, we show those PHDOM relations, which can be directly used in evaluation of the main energy-averaged characteristics of charge-exchange giant spin-multipole resonances in medium-heavy closed-shell parent nuclei. In Sect. III, these relations are employed for calculations of characteristics of charge-exchange giant spin-dipole GRs and overtones of these GRs in the parent nuclei mentioned above. Discussion of obtained results and a comparison with available experimental data are given in Sect. IV. Sect. V contains our summary and concluding remarks.

## II. Model relations

In spherical nuclei, IVGSMPR$^{(\mp)}$s are characterized by the quantum numbers of "transferred" isospin third projection $T_3 = \mp 1$ (related to excitations in the $\beta^{(\pm)}$ channels), total momentum $J$, spin $S = 1$ and, by neglecting tensor correlations, orbital momentum $L$ ($\vec{J} = \vec{L} + \vec{S}$, parity $\pi = (-1)^L$). Each $L>0$ IVGSMPR$^{(\mp)}$ has $2S+1$ $J$-components.

Since we include in the following analysis the GR double transition density, we start our consideration of main PHDOM relations (contrary to presentation of Ref. [4]), with the Bethe-Goldstone-type equation for the p-h Green functions (the effective p-h propagators) in the charge-exchange spin-flip channels, $\tilde{A}_S^{(\mp)}(x, x', \omega)$ ($x$ is a set of s-p coordinates, $\omega$ is the excitation energy) [2]. The mentioned equation contains, in particular, a p-h interaction responsible for formation of collective (p-h)–type excitations. Within PHDOM, this interaction is taken as Landau-Migdal forces, having the following spin-isospin part:

$$F_{L-M}^{s-is}(x_1, x_2) = G'(\vec{\tau}_1 \vec{\tau}_2)(\vec{\sigma}_1 \vec{\sigma}_2)\delta(\vec{r}_1 - \vec{r}_2) . \qquad (1)$$

For spherical nuclei, the effective p-h propagator might be expanded in terms of irreducible spin-tensor operators, $T_{JLSM}(\vec{n})$. By neglecting tensor

correlations and considering only the charge-exchange spin-flip p-h channels, one presents:

$$\tilde{A}_S^{(\mp)}(x,x',\omega) = (rr')^{-2} \sum_{JLM} \tilde{A}_{JLS}^{(\mp)}(r,r',\omega) T_{JLSM}(\vec{n}) T_{JLSM}^+(\vec{n}'). \quad (2)$$

Radial (two-dimensional) elements of these expansions obey the respective Bethe-Goldstone-type integral equations:

$$\tilde{A}_{JLS}^{(\mp)}(r,r',\omega) = A_{JLS}^{(\mp)}(r,r',\omega) + 2G' \int A_{JLS}^{(\mp)}(r,r_1,\omega) \tilde{A}_{JLS}^{(\mp)}(r_1,r',\omega) \frac{dr_1}{r_1^2}. \quad (3)$$

In these equations, the excitation energy of the $(Z \pm 1, N \mp 1)$ isobaric nuclei, $\omega = E_x + Q^{(\mp)}$, is counted off from the parent-nucleus $(Z, N)$ ground-state energy. Here, $Q^{(\mp)}$ are the differences of the ground-state energies of the respective isobaric and parent nuclei, and the excitation energy of the mentioned isobaric nuclei, $E_x$, is counted off from the ground-state energy of these nuclei. The key PHDOM quantity is the energy-averaged "free" p-h propagator $A(x, x', \omega)$ related to the model of non-interacting independently damping p-h excitations [2]. Listed in the Introduction, the main relaxation modes of these excitations are together taken into account in the expression for the "free" propagator. In Eq. (3), the quantities $(rr')^{-2} A_{JLS}^{(\mp)}(r,r',\omega)$ are radial elements of the expansion (like Eq. (2)) of the "free" p-h propagator taken in the charge-exchange spin-flip channels. To present rather cumbersome expressions for these radial elements, we employ the study of Ref. [12], where the expressions for the elements $4\pi(rr')^{-2}A_{000}^{(\mp)}(r,r',\omega)$, derived within PHDOM, are given in detail. The following substitution of the squared kinematic factors in these expressions,

$$(t_{(\pi)(\nu)}^{000})^2 = \frac{1}{4\pi}\delta_{(\pi)(\nu)}(2j_\pi + 1) \to (t_{(\pi)(\nu)}^{JLS})^2 = \frac{1}{2J+1}\langle(\pi)\|T_{JLS}\|(\nu)\rangle^2, \quad (4)$$

allows one to get the expressions for the elements $(rr')^{-2}A_{JLS}^{(\mp)}(r,r',\omega)$. These expressions contain: the occupation numbers $n_\mu$ for proton ($\mu = \pi$) and neutron ($\mu = \nu$) levels with $\mu$ being the set of single-particle quantum numbers $n_{r,\mu}$, $j_\mu$, $l_\mu$ (($\mu$) $= j_\mu$, $l_\mu$); the bound-state energies $\varepsilon_\mu$ and radial wave functions $r^{-1}\chi_\mu(r)$; and proton and neutron optical-model-like Green functions of the radial s-p Schrodinger equations, in which the mean field has an addition proportional to the strength of the p-h self-energy term responsible for the spreading effect, $(-iW(E_x) + P(E_x))$. The mentioned s-p radial Schrodinger equations determine also the proton and neutron optical-model-like radial continuum-state wave functions, $r^{-1}\chi_{\varepsilon>0,(\pi)}(r)$ and $r^{-1}\chi_{\varepsilon>0,(\nu)}(r)$, having the standing-wave asymptotic behavior and obeying the δ-function energy normalization in the limit $W = P = 0$. These wave functions are used below in describing strength functions of direct one-nucleon decay of IVGSMPR$^{(\mp)}$s.

The effective p-h propagators of Eqs. (2) and (3) determine, in particular, the double transition densities, having the radial (two-dimensional) elements,

$$\mathcal{R}_{JLS}^{(\mp)}(r,r',\omega) = -\frac{1}{\pi} Im \tilde{A}_{JLS}^{(\mp)}(r,r',\omega), \qquad (5)$$

as it follows from the spectral expansion of the effective propagator [2]. The radial elements of the double transition densities of Eq. (5) are related to the *J*-component of given IVGSMPR$^{(\mp)}$ and, in contrast to the cRPA limit ($W = P = 0$), cannot be factorized in terms of the respective radial elements of one-body transition densities. Strictly speaking, namely the energy-averaged double transition density (instead of the energy-averaged one-body transition density) should be used in an analysis of reaction cross sections of GR excitation [13].

Let $V_{JLSM}^{(\mp)}(x) = \tau^{(\mp)} V_L(r) T_{JLSM}(\vec{n})$ be s-p external fields (probing operators), leading to excitation of the *J*-component of IVGSMPR$^{(\mp)}$ of multipolarity $L$ ($\tau^{(\mp)}$ are the isobaric Pauli matrixes). The external-field radial part is taken as $V_L(r) = r^L$ for describing the respective main-tone GRs, and as $V_L^{ov}(r) = r^L(r^2 - \eta_L)$ for describing the related (first-order) overtone GRs. The parameter $\eta_L$ is found from the condition of minimal excitation of the main-tone GR in the $\beta^{(-)}$ channel by the respective "overtone" external field (see below).

The radial elements of the double transition densities of Eq. (5) determine the strength functions related to the above-mentioned external fields, as follows:

$$S_{JLS}^{(\mp)}(\omega) = -\frac{1}{\pi} Im \int V_L(r) \tilde{A}_{JLS}^{(\mp)}(r,r',\omega) V_L(r') dr dr'. \qquad (6)$$

The full and *J*-averaged strength functions, $S_L^{(\mp)}(\omega)$ and $\bar{S}_L^{(\mp)}(\omega)$, respectively,

$$S_L^{(\mp)}(\omega) = \sum_J (2J+1) S_{JLS}^{(\mp)}(\omega) = (2S+1)(2L+1)\bar{S}_L^{(\mp)}(\omega), \qquad (7)$$

are of practical interest. (Hereafter, the index $S = 1$ at full and $J$-averaged quantities is omitted). In a similar way, one can define $J$-averaged double transition densities, having the radial (two-dimensional) elements:

$$\bar{\mathcal{R}}_L^{(\mp)}(r,r',\omega) = \frac{1}{(2S+1)(2L+1)} \sum_J (2J+1)\, \mathcal{R}_{JLS}^{(\mp)}(r,r',\omega). \qquad (8)$$

These quantities determine, in particular, the $J$-averaged strength functions of Eq. (7):

$$\bar{S}_L^{(\mp)}(\omega) = \int V_L(r)\, \bar{\mathcal{R}}_L^{(\mp)}(r,r',\omega) V_L(r') dr dr'. \qquad (9)$$

Within PHDOM, the most of GR characteristics might be expressed in terms the effective field introduced in nuclear physics by Migdal [14]. The effective fields, $\tilde{V}_{JLSM}^{(\mp)}(x,\omega)$, have the same isobaric and, in neglecting tensor correlations, spin-angular dependence, as the above-considered external fields have [6]. Within PHDOM, the effective-field radial parts, $\tilde{V}_{JLS}^{(\mp)}(r,\omega)$, are defined by the integral relations:

$$\int \tilde{A}_{JLS}^{(\mp)}(r,r',\omega)V_L(r')dr' = \int A_{JLS}^{(\mp)}(r,r',\omega)\tilde{V}_{JLS}^{(\mp)}(r',\omega)dr'. \qquad (10)$$

These relations together with PHDOM basic equations for elements of the effective p-h propagator (Eq. (3)) allow one to get the equations for the effective-field radial parts:

$$\tilde{V}_{JLS}^{(\mp)}(r,\omega) = V_L(r) + \frac{2G'}{r^2}\int A_{JLS}^{(\mp)}(r,r',\omega)\tilde{V}_{JLS}^{(\mp)}(r',\omega)dr'. \qquad (11)$$

These equations are, obviously, simpler than Eq. (3).

The effective-field radial parts of Eq. (11) determine most of the main characteristics of IVGSMPR$^{(\mp)}$ $J$-components. The expressions for these characteristics, related to the respective external field and considered in a wide excitation-energy interval, are given below.

(i) The expressions for the strength functions $S_{JLS}^{(\mp)}(\omega)$ follow from Eqs. (6) and (10):

$$S_{JLS}^{(\mp)}(\omega) = -\frac{1}{\pi}Im \int V_L(r)\, A_{JLS}^{(\mp)}(r,r',\omega)\tilde{V}_{JLS}^{(\mp)}(r',\omega)drdr'. \qquad (12)$$

(ii) The expressions for the radial (one-dimensional) elements of the projected (one-body) transition densities, $\rho_{JLS}^{(\mp)}(r,\omega)$, follow from the definition of these quantities [13],

$$\rho_{JLS}^{(\mp)}(r,\omega) = \int \mathcal{R}_{JLS}^{(\mp)}(r,r',\omega)V_L(r')dr'/(S_{JLS}^{(\mp)}(\omega))^{1/2}, \qquad (13)$$

and the use of Eqs. (6), (10), and (11):

$$r^{-2}\rho_{JLS}^{(\mp)}(r,\omega) = -\frac{1}{\pi}Im\tilde{V}_{JLS}^{(\mp)}(r,\omega)/2G'\left[S_{JLS}^{(\mp)}(\omega)\right]^{1/2}. \qquad (14)$$

Similarly to the definition of Eq. (13), one can define radial elements of $J$-averaged projected transition densities:

$$\bar{\rho}_L^{(\mp)}(r,\omega) = \int \bar{\mathcal{R}}_L^{(\mp)}(r,r',\omega)V_L(r')dr'/\left[\bar{S}_L^{(\mp)}(\omega)\right]^{1/2}. \qquad (15)$$

These quantities might be expressed in terms of the respective $J$-components according to definitions of Eqs. (8) and (15):

$$\bar{\rho}_L^{(\mp)}(r,\omega) = \frac{1}{(2S+1)(2L+1)}\sum_J (2J+1)\rho_{JLS}^{(\mp)}(r,\omega)\left[\frac{S_{JLS}^{(\mp)}(\omega)}{\bar{S}_L^{(\mp)}(\omega)}\right]^{1/2}. \qquad (16)$$

(iii) The expressions for the doubly-partial and partial strength functions (differential probabilities) of IVGSMPR$^{(-)}$ direct one-proton decay accompanied by population of the product-nucleus neutron-hole states $\nu^{-1}$, $S_{JLS,(\pi)\nu}^{(-),\uparrow}$ and $S_{JLS,\nu}^{(-),\uparrow}$, respectively, are the following [2, 15]:

$$S^{(-),\uparrow}_{JLS,(\pi)\nu}(\omega) = n_\nu (t^{JLS}_{(\pi)(\nu)})^2 \left| \int \chi^*_{\varepsilon=\varepsilon_\nu+\omega,(\pi)}(r) \tilde{V}^{(-)}_{JLS}(r,\omega) \chi_\nu(r) dr \right|^2 \quad (17)$$

and

$$S^{(-),\uparrow}_{JLS,\nu} = \sum_{(\pi)} S^{(-),\uparrow}_{JLS,(\pi)\nu}(\omega). \quad (18)$$

The respective expressions for the direct one-neutron-decay strength functions related to the J-component of IVGSMPR$^{(+)}$ follow from Eqs. (17) and (18) after substitutions: (-) → (+), ν ↔ π. (These expressions are abbreviated below as Eqs. (17') and (18')).

We now provide comments to the above-given expressions for main characteristics of IVGSMPR$^{(\mp)}$s.

1) The strength functions $S^{(\mp)}_{JLS}(E_x)$ obey the non-energy-weighted sum rule (NEWSR$_L$), which is independent of $J$:

$$\text{NEWSR}_L = \int S^{(-)}_{JLS}(E_x) dE_x - \int S^{(+)}_{JLS}(E_x) dE_x = \int V_L^2(r) n^{(-)}(r) r^2 dr. \quad (19)$$

Here, $n^{(-)}(r)$ is the neutron-excess density in the parent nucleus. This relation is also valid in cRPA. To verify the strength function calculations, it is reasonable to compare with unity the fraction parameter $x_L^c$ defined for a large cut-off excitation energy $E_x^c$:

$$x_L^c = \left( \int_0^{E_x^c} S^{(-)}_{JLS}(E_x) dE_x - \int_0^{E_x^c} S^{(+)}_{JLS}(E_x) dE_x \right) / \text{NEWSR}_L. \quad (20)$$

The *J*-averaged strength functions of Eq. (7) obey the sum rule of Eq. (19) and determine the *J*-averaged fraction parameter, $\bar{x}_L^c = \bar{x}_L^{(-),c} - \bar{x}_L^{(+),c}$, similarly to Eq. (20). In the same way, one can consider the fraction parameters $\bar{x}_L^{(\mp)}(\delta_{12}^{(\mp)}) = \int_{E_{x,1}}^{E_{x,2}} \bar{S}_L^{(\mp)}(E_x) dE_x / \text{NEWSR}_L$ defined for given excitation-energy intervals $\delta_{12}^{(\mp)} = E_{x,1} \div E_{x,2}$.

2) The above-given expressions related to characteristics of main-tone GRs are also valid for overtone GRs after the substitution: $V_L(r) \to V_L^{ov}(r)$. The parameter $\eta_L$ in the definition of the radial part $V_L^{ov}(r)$ of the probing operator, leading to excitation of related IVGSMPR$^{(\mp)}$2, can be found from the condition of minimal excitation of IVGSMPR$^{(-)}$ by the respective "overtone" operator [4]: $min \int S_L^{(-),ov}(E_x) dE_x$, where integration is performed over the main-tone GR region. This region is placed at the low-energy distant "tail" of the overtone-GR full strength function.

3) Calculations of the (one-dimensional) radial elements of the projected one-body transition densities, $\rho_{JLS}^{(\mp)}(r,\omega)$, performed in accordance with Eq. (14) might be verified by the relations,

$$S_{JLS}^{(\mp)}(\omega) = \left( \int V_L(r) \rho_{JLS}^{(\mp)}(r,\omega) \, dr \right)^2, \tag{21}$$

which are following from Eqs. (13), (5) and (6). As follows from the comparison of these relations with Eqs. (5) and (6), only in evaluating the strength functions $S_{JLS}^{(\mp)}(\omega)$ the double transition densities might be approximated by the product of respective (i.e., related to the same external field) projected one-body transition densities and therefor might be considered as the projected double transition density:

$$\mathcal{R}_{JLS}^{(\mp),pr}(r,r',\omega) = \rho_{JLS}^{(\mp)}(r,\omega)\rho_{JLS}^{(\mp)}(r',\omega). \qquad (22)$$

Within cRPA, factorization of the double transition density takes place independently of the external field.

4) The full strength function of IVGSMPR$^{(-)}$ direct one-proton decay into channel $\nu$, $S_{L,\nu}^{(-),\uparrow}(\omega)$, is defined via the respective $J$-components of Eq. (18) by a relation similar to Eq. (7). This strength function together with the full strength function of Eq. (7) determines the partial branching ratio of the above-mentioned decay from a given excitation-energy interval:

$$b_{L,\nu}^{(-),\uparrow}\left(\delta_{12}^{(-)}\right) = \int_{\delta_{12}^{(-)}} S_{L,\nu}^{(-),\uparrow}(\omega)d\omega \,/\, \int_{\delta_{12}^{(-)}} S_L^{(-)}(\omega)d\omega. \qquad (23)$$

Substitutions (-) → (+), $\nu \to \pi$ in Eq. (23) together with Eqs. (17') and (18') lead to an expression (abbreviated below as Eq. (23')) for the partial branching ratio of direct one-neutron decay of IVGSMPR$^{(+)}$ into channel $\pi$. Respectively, the expressions for total branching ratios of direct one-nucleon decay of IVGSMPR$^{(\mp)}$ follow from Eqs. (23) and (23'):

$$b_{L,tot}^{(-),\uparrow}\left(\delta_{12}^{(-)}\right)=\sum_{\nu}b_{L,\nu}^{(-),\uparrow}\left(\delta_{12}^{(-)}\right),\ b_{L,tot}^{(+),\uparrow}\left(\delta_{12}^{(+)}\right)=\sum_{\pi}b_{L,\pi}^{(+),\uparrow}\left(\delta_{12}^{(+)}\right). \quad (24)$$

In neglecting pre-equilibrium decay, the differences of full strength functions, $S_L^{(-)}(\omega)-\sum_{\nu}S_{L,\nu}^{(-),\uparrow}(\omega)=S_L^{(-),\downarrow}(\omega)$ and $S_L^{(+)}(\omega)-\sum_{\pi}S_{L,\pi}^{(+),\uparrow}(\omega)=S_L^{(+),\downarrow}(\omega)$, can be considered as the statistical-decay strength functions of IVGSMPR$^{(\mp)}$. Within cRPA, these strength functions go to zero (the unitary condition for one-nucleon decay strength functions). Respectively, the total branching ratios of direct one-nucleon decay determine the branching ratios of statistical decay of IVGSMPR$^{(\mp)}$:

$$b_L^{(\mp),\downarrow}=1-b_{L,tot}^{(\mp),\uparrow}. \quad (25)$$

These relations are following from Eqs. (23), (23') and (24) and the above-given definition of the statistical-decay strength functions. Comparison with unity of branching ratios $b_{L,tot}^{(\mp),\uparrow}$ of Eqs. (24) (or $b_L^{(\mp),\downarrow}$ of Eqs. (25)) allows one to estimate the contribution of the spreading effect to formation of respective GRs.

5) In contrast with the GR strength function and transition density, which are the characteristics of collective motion, the probabilities (or branching ratios) of GR direct one-nucleon decays are the characteristics of interplay of collective and s-p motions. To make more adequate comparison of evaluated by Eqs. (17), (17'),

(18) and (18') the direct one-nucleon decay branching ratios of Eqs. (23) and (23') with respective experimental data, we, following Ref. [7], recalculate these branching ratios. The recalculating procedure takes into account: (i) a more complicated structure of product-nuclei one-hole states $\mu^{-1}$ populated after decay (by the use of the respective spectroscopic factor $SF_\mu$); (ii) sensitivity of the direct-decay characteristics to potential-barrier penetrability for escaped nucleons (by the use of the s-p optical-model transmission coefficients $T_\lambda(\varepsilon)$). For instance, the expression for, recalculated in accordance with Eq. (17), the doubly-partial strength function of direct one-proton decay of the IVGSMPR$^{(-)}$ $J$-component has the form:

$$\check{S}_{JLS,(\pi)\nu}^{(-),\uparrow}(\omega) = SF_\nu S_{JLS,(\pi)\nu}^{(-),\uparrow}(\omega) T_\pi(\varepsilon = \varepsilon_\nu + \omega + \Delta\varepsilon)/T_\pi(\varepsilon = \varepsilon_\nu + \omega). \quad (26)$$

Here, $T_\pi = 1 - \exp[-4\eta_\pi]$, where $\eta_\pi$ is the imaginary part of the s-p optical-model scattering phase-shift, and $\Delta\varepsilon = \Delta\varepsilon_\nu + \Delta\omega_{m_J}$ with $\Delta\varepsilon_\nu = \varepsilon_\nu^{exp} - \varepsilon_\nu$ and $\Delta\omega_{m_J} = \omega_{m_J}^{exp} - \omega_{m_J}$ being differences of the experimental and calculated energies of the neutron-hole state and of the GR maximum, respectively.

The last comment concludes presentation of the PHDOM-based relations, allowing evaluate main characteristics of IVGSMPR$^{(\mp)}$s in medium-heavy closed-shell parent nuclei. In the next Section, these relations are used in applying to IVGSDR$^{(\mp)}$ and IVGSDR$^{(\mp)}$2 ($J^\pi = 0^-, 1^-$ and $2^-, L = S = 1$). Here, we note, that the sum rule of Eq. (19) for the $L = 1$ main-tone GRs,

$$NEWSR_{L=1} = \frac{1}{4\pi} \int r^2 n^{(-)}(r) d\vec{r} = \frac{1}{4\pi}(N\langle r^2\rangle_n - Z\langle r^2\rangle_p), \tag{27}$$

is related to the neutron skin of the respective parent nucleus. For this reason, experimental studies of the strength functions of IVGSDR$^{(\mp)}$ are the object of special interest.

### III. Properties of IVGSDR$^{(\mp)}$ and their overtones

The following input quantities are used in employing relations of Sect. II for evaluation of main characteristics of IVGSDR$^{(\mp)}$ and their overtones in the $^{48}$Ca, $^{90}$Zr, $^{132}$Sn, and $^{208}$Pb parent nuclei: (i) the realistic (Woods-Saxon type) phenomenological partially self-consistent mean field described in details in Ref. [3]; (ii) Landau-Migdal forces of Eq. (1); (iii) the imaginary part $W(E_x)$ of the strength of the energy-averaged p-h self-energy term responsible for the spreading effect (the real part, $P(E_x)$, is determined by the imaginary part via the proper dispersive relationship). The values of mean-field parameters, the Landau-Migdal parameter $g' = G'/300$ MeV fm$^3$, and "spreading" parameters (parameters of the energy dependence of $W(E_x)$) have been specified in Ref. [4], where the PHDOM-based description of Gamow-Teller and charge-exchange giant spin-monopole resonances in the parent nuclei under consideration has been proposed. These parameters (listed in Table I of Ref. [4]) are used in evaluation of characteristics of the considered dipole resonances.

We start presenting the calculation results with the evaluated strength functions $S_J^{(\mp)}(E_x)$ and $\bar{S}^{(\mp)}(E_x)$ of IVGSDR$^{(\mp)}$ in parent nuclei under consideration (hereafter, the indexes $L = S = 1$ at characteristics of spin-dipole GRs are omitted). Evaluated by Eq. (12) the $J$-component strength functions of IVGSDR$^{(-)}$ in parent nuclei under consideration are shown in Fig. 1 together with the $J$-averaged strength function of Eq. (7). This presentation is specified in Fig.2, where the evaluated J-component and full strength functions of IVGSDR$^{(-)}$ in $^{208}$Bi, $(2J+1)S_J^{(-)}(\omega)$, $S^{(-)}(\omega)$ and $S^{(-)}(\omega)$, are shown in a comparison with respective quantities deduced from experimental data of Ref. [16]. For a comparison with experimental data, the evaluated $J$-averaged strength functions are approximated by the Lorentz-type energy dependence and, also, compared with the respective cRPA limit (Fig. 3). Similarly to the results shown in Figs. 1 and 3, the evaluated strength functions of IVGSDR$^{(+)}$ in the $^{48}$Ca and $^{90}$Zr parent nuclei are shown in Fig. 4. In these nuclei (in contrast with the $^{132}$Sn and $^{208}$Pb parent nuclei) the Pauli blocking-effect decreases the IVGSDR$^{(+)}$ strength without a marked change of GR structure. In heavy nuclei, having a large neutron excess, this effect reduces IVGSDR$^{(+)}$ up to a few resonances related to weak low-energy (p-h)-type excitations. This statement is illustrated by Fig. 5, where the evaluated $J$-averaged strength functions of IVGSDR$^{(+)}$ in $^{132}$Sn and $^{208}$Pb parent nuclei are shown. . The small total relative strength of these resonances is discussed in Sect IV. As for overtone GRs, we show in Figs. 5 and 6 (similarly to Figs. 3 and 4) only the evaluated *J*-averaged strength functions of IVGSDR$^{(\mp)}$2 in parent nuclei under consideration.

From the evaluated *J*-averaged strength functions one can deduce the following parameters of IVGSDR$^{(\mp)}$ and IVGSDR$^{(\mp)}$2 in parent nuclei under consideration: (i) the fraction parameters $\bar{x}^c$ and $\bar{x}^{(\mp),c}$ evaluated according to Eqs. (7) and (20) for the cut-off excitation-energy $E_x^c = 80$ MeV; (ii) the energy $E_{x,m}^{(\mp)}$ and total width $\Gamma^{(\mp)}$ of the respective strength-function maximum approximated by the Lorentz-type energy dependence; (iii) large enough excitation-energy intervals $\delta_{12}^{(\mp)} = E_{x,1} \div E_{x,2}$, which include the respective main strength-function maximum, related to these intervals the relative strengths $\bar{x}_{12}^{(\mp)}/x_c^{(\mp)}$, mean excitation energies $E_{x,mean}^{(\mp)}$ (defined as the ratio of the first to zero moments), and energy dispersions $\sigma^{(\mp)}$. The above-listed parameters are given in Table II for IVGSDR$^{(-)}$ in parent nuclei under consideration and in Table III for IVGSDR$^{(+)}$ in the $^{48}$Ca, $^{90}$Zr parent nuclei. In Table IV, the reduced list of parameters evaluated for IVGSDR$^{(\mp)}$2 in parent nuclei under consideration is given. The values of parameter $\eta$ in definition of the "overtone" external-field radial part are also given in Table IV.

Table I. The model parameters specified in Ref. [4] and used for evaluation of characteristics of IVGSDR$^{(\mp)}$ and IVGSDR$^{(\mp)}$2 in parent nuclei under consideration. The mean-field parameter $r_0 = 1.21$ fm and spreading parameter $\Delta = 3$ MeV are taken as universal quantities [4].

| Nucleus | $U_0$, (MeV) | $U_{ls}$, (MeV fm$^2$) | $a$, (fm) | $f'$ | $g'$ | $\alpha$, (MeV$^{-1}$) | $B$, (MeV) |
|---|---|---|---|---|---|---|---|
| $^{48}$Ca | 54.34 | 32.09 | 0.58 | 1.13 | 0.85 | 0.25 | 5.34 |
| $^{90}$Zr | 55.06 | 34.93 | 0.61 | 1.05 | 0.68 | 0.51 | 5.24 |
| $^{132}$Sn | 55.53 | 35.98 | 0.63 | 1.00 | 0.71 | 0.26 | 5.84 |
| $^{208}$Pb | 55.74 | 33.35 | 0.63 | 0.98 | 0.73 | 0.24 | 5.12 |

Table II. Evaluated within PHDOM the parameters of IVGSDR$^{(-)}$ in parent nuclei under consideration. (Notations are given in the text).

| Nucleus | $\bar{x}^c$, % | $\bar{x}^{(-),c}$, % | $E_{x,m}^{(-)}$, MeV | $\Gamma^{(-)}$, MeV | $\delta_{12}^{(-)}$, MeV | $\bar{x}_{12}^{(-)}/\bar{x}^{(-),c}$ | $E_{x,mean}^{(-)}$, MeV | $\sigma^{(-)}$, MeV | $b_{L,tot}^{(-),\uparrow}$, % |
|---|---|---|---|---|---|---|---|---|---|
| $^{48}$Ca | 106.4 | 128.3 | 19.2 | 13.6 | 5.7 ÷ 32.8 | 0.90 | 22.1 | 8.8 | 40.3 |
| $^{90}$Zr | 108.3 | 145.0 | 14.8 | 16.9 | 0.0 ÷ 31.6 | 0.89 | 20.8 | 10.4 | 16.3 |
| $^{132}$Sn | 96.6 | 103.5 | 22.1 | 14.2 | 7.9 ÷ 36.4 | 0.92 | 24.6 | 9.0 | 15.5 |
| $^{208}$Pb | 98.4 | 104.5 | 21.2 | 12.3 | 8.9 ÷ 33.5 | 0.90 | 23.2 | 7.9 | 15.8 |

Table III. The same as in Table II, but for IVGSDR$^{(+)}$ in $^{48}$Ca and $^{90}$Zr parent nuclei.

| Nucleus | $\bar{x}^c$, % | $\bar{x}^{(+),c}$, % | $E_{x,m}^{(+)}$, MeV | $\Gamma^{(+)}$, MeV | $\delta_{12}^{(+)}$, MeV | $\bar{x}_{12}^{(+)}/\bar{x}^{(+),c}$ | $E_{x,mean}^{(+)}$, MeV | $\sigma^{(+)}$, MeV | $b_{L,tot}^{(+),\uparrow}$, % |
|---|---|---|---|---|---|---|---|---|---|
| $^{48}$Ca | 106.4 | 21.9 | 10.3 | 5.8 | 4.6 ÷ 16.1 | 0.74 | 12.0 | 4.0 | 16.0 |
| $^{90}$Zr | 108.3 | 36.8 | 9.9 | 7.9 | 1.9 ÷ 17.8 | 0.77 | 13.4 | 4.9 | 2.8 |

Table IV. Evaluated within PHDOM the parameters of IVGSDR$^{(\mp)}$2 in parent nuclei under consideration.

| Nucleus | $\eta, fm^2$ | $\bar{x}^c$, % | $\bar{x}^{(-),c}$, % | $E_{x,m}^{(-)}$, MeV | $\Gamma^{(-)}$, MeV | $\bar{x}^{(+),c}$, % | $E_{x,m}^{(+)}$, MeV | $\Gamma^{(+)}$, MeV |
|---|---|---|---|---|---|---|---|---|
| $^{48}$Ca | 23.5 | 98.7 | 144.2 | 31.7 | 26.4 | 45.5 | 24.4 | 22.9 |
| $^{90}$Zr | 30.2 | 96.4 | 212.6 | 33.1 | 27.2 | 116.1 | 21.3 | 29.1 |
| $^{132}$Sn | 38.6 | 98.1 | 120.1 | 36.2 | 22.5 | 22.0 | 15.9 | 16.2 |
| $^{208}$Pb | 48.6 | 98.3 | 125.2 | 37.5 | 19.6 | 26.9 | 16.4 | 12.8 |

A large volume of calculation results concerned with the projected (one-body) and double transition densities of IVGSDR$^{(\mp)}$ and IVGSDR$^{(-)}$2 in parent nuclei under consideration are illustrated by a few examples. In particular, for the mentioned nuclei we show in Figs. 8 and 9 the radial elements of the projected transition densities $\rho_{J=1}^{(-)}(r, E_{x,m_1})$ calculated by Eq. (14) at the maximum of the strength functions of, respectively, $1^{(-)}$-IVGSDR$^{(-)}$ and its overtone. Radial elements of various projected transition densities, $\rho_J^{(-)}(r, E_{x,m_J})$, $\bar{\rho}^{(-)}(r, E_{x,m})$, evaluated according to Eqs. (14) and (15), are compared in Fig. 10, where these elements are shown for IVGSDR$^{(-)}$ in $^{208}$Bi taken as an example. To imagine the difference between the real and projected double transition densities (Eq. (5) and Eq. (22), respectively), we consider these densities for $1^{(-)}$-IVGSDR$^{(-)}$ in $^{208}$Bi also taken as an example. In Fig. 11, the radial elements of these densities evaluated at $E_x = E_{x,m_1}$ are compared along two lines: $r = r'$ and $r + r' = 2r_m$ ($r_m$ is related to the maximum of $\rho_{J=1}^{(-)}(r, E_{x,m_1})$ radial dependence shown in Fig. 8).

Presenting the branching ratios of direct one-nucleon decay of IVGSDR$^{(\mp)}$ in nuclei under consideration evaluated within PHDOM is started with the total branching ratios $b_{tot}^{(\mp),\uparrow}$ of Eq. (24) (Table II and Table III). Given in these Tables the branching ratios are evaluated for the chosen excitation-energy intervals $\delta_{12}^{(\mp)}$. Direct one-proton decay of IVGSDR$^{(-)}$ in $^{208}$Bi is the object of special study, because in such case the experimental data concerned with partial branching ratios for a few decay-channels are available [17]. In Table V, the partial branching ratios

$b_\nu^{(-),\uparrow}(\delta_{12}^{(-)})$ evaluated according to Eqs. (17), (18) and (23) for mentioned decay-channels and the recalculated branching ratios $\check{b}_\nu^{(-),\uparrow}(\delta_{12}^{(-)})$ of Eqs. (26), (18) and (23) are given together with experimental spectroscopic factors, $SF_\nu^{exp}$, experimental and calculated s-p energies, $\varepsilon_\nu^{exp}$ and $\varepsilon_\nu$, respectively. The values $\Delta\omega_{m_J} = 0$ are used in calculations performed with the use of Eq. (26), because the experimental and calculated energies $E_{x,m}^{(-)}$ of the considered GR are in close agreement (see below). Since the similar experimental data for Gamow-Teller resonance (GTR) in $^{208}$Bi are also known [18], we evaluate the respective partial branching ratios of this resonance, using the same calculation scheme and the same set of model parameters. The obtained results can be considered as a reasonable supplement to our recent study of Ref. [4] (see Introduction). Two sets of partial branching ratios calculated for considered GRs are compared with respective experimental data of Refs. [17, 18] (Table V).

Table V. The partial branching ratios of direct one-proton decay of GTR and IVGSDR$^{(-)}$ in $^{208}$Bi evaluated within PHDOM for a few decay-channels (notations are given in the text). The experimental data are taken from Refs. [17-20]. The related excitation-energy intervals considered are given in Table I of Ref. [4] and Table II of this work.

| $\nu^{-1}$ | $SF_\nu^{exp}$ [19] | $\varepsilon_\nu$, MeV | $\varepsilon_\nu^{exp}$, MeV [20] | GTR | | | IVGSDR$^{(-)}$ | | |
|---|---|---|---|---|---|---|---|---|---|
| | | | | $b_\nu^{(-),\uparrow}$, % | $\check{b}_\nu^{(-),\uparrow}$, % | $b_{\nu,exp}^{(-),\uparrow}$, % [18] | $b_\nu^{(-),\uparrow}$, % | $\check{b}_\nu^{(-),\uparrow}$, % | $b_{\nu,exp}^{(-),\uparrow}$, % [17] |
| $3p_{1/2}$ | 1.0 | 7.23 | 7.37 | 1.15 | 1.05 | 1.8 ± 0.5 | 1.20 | 1.18 | 0.95 ± 0.28 |
| $2f_{5/2}$ | 0.98 | 8.08 | 7.94 | 2.50 | 2.75 | Incl. in $3p_{3/2}$ | 2.49 | 2.49 | 2.10 ± 0.61 |
| $3p_{3/2}$ | 1.0 | 8.3 | 8.27 | 1.05 | 1.07 | 2.7 ± 0.6 | 2.42 | 2.42 | 2.79 ± 0.81 |
| $1i_{13/2}$ | 0.91 | 9.22 | 9.00 | 0.01 | 0.01 | 0.2 ± 0.2 | 3.51 | 3.19 | 3.41 ± 0.98 |
| $2f_{7/2}$ | 0.7 | 10.8 | 10.07 | 0.08 | 0.13 | 0.4 ± 0.2 | 2.78 | 2.24 | 3.14 ± 0.91 |
| $1h_{9/2}$ | 0.61 | 11.51 | 10.78 | 0.02 | 0.03 | - | 1.11 | 0.82 | 0.97 ± 0.27 |
| Sum | | | | 4.81 | 5.05 | 4.9 ± 1.3 | 13.50 | 12.34 | 13.4 ± 3.9 |

## IV. Discussion of results

Discussion of calculation results is started with the strength functions and related parameters of IVGSDR$^{(\mp)}$ in parent nuclei under consideration (Figs. 1-5, Table II and Table III). (i) The mentioned strength functions exhaust well enough the non-energy-weighted sum rule of Eq. (19). The evaluated $J$-averaged fraction parameters $\bar{x}_L^c$ of Eq. (20) are close to unity (Table II). Relatively small deviations from unity might be related to neglecting by RPA ground-state correlations ("core-polarization" effect) in evaluation of the neutron-excess density employed in evaluation of $NEWSR_{L=1}$ accordingly Eq. (19). An attempt to take this effect into account within cRPA has been undertaken in Ref. [21], but not completed. (ii) Being the ratio of integral strengths of IVGSDR$^{(+)}$ and IVGSDR$^{(-)}$, the ratio of fraction parameters $\bar{x}_L^{(+)c}/\bar{x}_L^{(-)c}$ depends essentially on the Pauli blocking-effect and reaches the small values of 0.07 and 0.06 for the $^{132}$Sn and $^{208}$Pb parent nuclei, respectively. This ratio is not-too-small for the $^{48}$Ca and $^{90}$Zr parent nuclei (Table II and Table III). Strong suppression of the IVGSDR$^{(+)}$ strength due to Pauli-blocking effect takes place in heavy nuclei. In these nuclei, the inter-shell distance (the mean energy of s-p dipole transitions) is comparable with the mean energy-difference of s-p proton and neutron levels, having the same quantum numbers (approximately this difference is proportional to N-Z). (iii) The energy dependence of strength functions $S_J^{(-)}(E_x)$ related to the $J=0$ and $J=1$ components of IVGSDR$^{(-)}$ exhibits the well-formed resonance, having weak pigmy resonance(s) at its low-energy part (Fig. 1). The energy dependence of strength function $S_{J=2}^{(-)}(E_x)$ exhibits a two-bump resonance, having a rather strong pygmy resonance. For this reason (due to large statistical factor related to the ($J=2$)-component in Eq. (7)), the $J$-averaged strength function of IVGSDR$^{(-)}$ also exhibits the two-bump energy dependence, which is purely described by the Lorentz-type energy dependence (Figs. 1 and 3). Additional broadening of the main maximum of $\bar{S}^{(-)}(E_x)$ strength function, appearing, within the model, due to difference of the energy of $S_J^{(-)}(E_x)$ maxima and to the two-bump structure of $S_{J=2}^{(-)}(E_x)$, might be considered as "second-order" Landau damping of IVGSDR$^{(-)}$. (iv) As it follows from a comparison of strength functions $\bar{S}^{(-)}$ evaluated within cRPA and PHDOM (Fig. 3), the spreading effect gives the main contribution to formation of IVGSDR$^{(-)}$ in the $^{90}$Zr, $^{132}$Sn, and $^{208}$Pb parent nuclei. The total branching ratio of direct one-proton decay of these GRs (evaluated by Eq. (24)) is about 16% (Table II). Only for $^{48}$Ca, contributions of Landau damping + s-p continuum and spreading effect are comparable (the mentioned branching ratio is about 40% (Table II)). (v) As mentioned above, Pauli blocking-effect on properties of IVGSDR$^{(+)}$ in the $^{48}$Ca and $^{90}$Zr parent nuclei is not-too-large. For this reason, the energy dependence of strength functions $S_J^{(+)}(E_x)$ and $\bar{S}^{(+)}(E_x)$ is similar to that for the respective strength functions of

IVGSDR$^{(-)}$ in mentioned parent nuclei (Fig. 4 and Figs. 1 and 3). Due to Pauli blocking-effect, the strength of IVGSDR$^{(+)}$ in the $^{132}$Sn and $^{208}$Pb parent nuclei is strongly suppressed (p. (ii)). As a result, the respective strength functions $\bar{S}^{(+)}(E_x)$ exhibit only a few weak low-energy resonances (Fig. 5) and are essentially different from the related strength functions $\bar{S}^{(-)}(E_x)$ (Fig. 3). (vi) Parameters of considered IVGSDR$^{(\mp)}$ deduced from the evaluated $J$-averaged strength functions discussed above are listed in Table II and Table III (notations are given earlier in this Section). The respective experimental data are scant. We note here the experimental values of energy $E_{x,m}^{(-)} = 21.1 \pm 0.8$ MeV and total width $\Gamma^{(-)} = 8.4 \pm 1.7$ MeV deduced in Ref. [18] from the $^{208}$Pb($^3$He,t)-reaction cross section. The respective calculated values (Table II) are in a reasonable agreement with these experimental data. (vii) The $J$-component and full strength functions deduced from an analysis of $^{208}$Pb $(\vec{p},\vec{n})$-reaction cross sections [16] are shown in Fig. 2 in a comparison with the respective quantities evaluated within PHDOM. As follows from this comparison the full strength function is reasonably described. Here, we note a possibility to describe, within the model, contribution of tensor correlations to formation of IVGSMPR$^{(\mp)}$. As follows from preliminary study of Ref. [6], an adequate description of the spreading effect is needed to describe details of strength distribution. From this point of view, the RPA-based studies of tensor correlations undertaken in Refs. [16, 22] seem to us not completed.

As mentioned in the Introduction, PHDOM is an extension of cRPA. For this reason, there is a possibility to describe within the model properties of high-energy GRs. In the present study, this possibility is realized for a description of the overtone GRs, IVGSDR$^{(\mp)}$2, in the parent nuclei under consideration. Evaluated within the model the *J*-averaged strength functions of IVGSDR$^{(-)}$2 and IVGSDR$^{(+)}$2 (Figs. 6 and 7, respectively) exhaust well enough the related NEWSR: the respective fraction parameters $\bar{x}^c$ are found to be close to unity (Table IV). The strength functions exhibit a broad resonance, which is formed mainly due to Landau damping + s-p continuum. Parameters of this resonance, the energy of the strength-function maximum and total width (Table IV), are typical for high-energy GRs. These parameters, as a majority of the IVGSDR$^{(\mp)}$ parameters, evaluated within PHDOM, can be considered as a prediction for future experimental studies.

In addition to strength distributions, transition densities also belong to main characteristics of GRs. As examples, we show in Figs. 8 and 9 the radial elements of projected (one-dimensional) transition densities, $\rho_{J=1}^{(-)}(r, E_{x,m_1})$ and $\rho_{J=1}^{(-),ov}(r, E_{x,m_1})$, evaluated at maximum of the (*J*=1)-component strength function of, respectively, IVGSDR$^{(-)}$ and IVGSDR$^{(-)}$2 in parent nuclei under consideration. As expected for well-formed GRs, these transition densities exhibit, respectively, node-less and one-node radial dependence. A possibility of approximate factorization of the energy-averaged double transition density might be illustrated by comparison of this quantity with its projected partner (Eq. (22)). Considering (*J*=1)- IVGSDR$^{(-)}$ in $^{208}$Bi as an example, we compare in Figs. 11a and 11b, respectively, diagonal and "anti-diagonal" parts of radial elements of these transition densities evaluated at maximum of strength function $S_{J=1}^{(-)}(E_x)$ (Fig. 1).

Proximity of real and projected double transition densities, considered at GR maximum, makes reasonable to employ the projected (one-body) transition density in an analysis of cross sections of GR excitation (see, also, Ref. [13]). The example of evaluated $J$-dependent and $J$-averaged radial elements of the projected transition density of IVGSDR$^{(-)}$ in $^{208}$Bi is given in Fig. 10.

Ability to describe probabilities (or branching ratios) of GR direct one-nucleon decay is a unique feature of PHDOM. In Table V, we show partial branching ratios calculated within the model and then properly specified for a few channels of direct one-proton decay of GTR and IVGSDR$^{(-)}$ in $^{208}$Bi. Calculation results are in reasonable agreement with respective experimental data. The use of specification procedure does not change significantly the description of experimental branching ratios.

V. Summary and concluding remarks

In this work, we present the method for a theoretical investigation of main properties of charge-exchange giant spin-multipole resonances in medium-heavy closed-shell parent nuclei. Being based on the properly adapted semimicroscopic particle-hole dispersive optical model, the method is applied to the description of main characteristics and parameters of charge-exchange giant spin-dipole resonances and their overtones in the $^{48}$Ca, $^{90}$Zr, $^{132}$Sn, and $^{208}$Pb parent nuclei. Giant-resonance strength distributions, transition densities, probabilities (branching ratios) of direct one-nucleon decay are evaluated with the use of model parameters taken from similar study of main properties of Gamow-Teller and charge-exchange giant spin-monopole resonances in the same parent nuclei. Calculation results are in reasonable agreement with experimental data, which are few in number, so that most of obtained results might be considered as predictions. Presented in this work and obtained early results support the statement that the employed model is useful tool for theoretical investigation of properties of various giant resonances in medium-heavy closed-shell nuclei. Implementation of the properly extended model to taking into account the effect of tensor correlations on formation of giant resonances is in order.

## Acknowledgments.

The authors are thankful to M. Fujiwara for viewing the manuscript and valuable remarks.

This work is partially supported by Program "Priority 2030" for National Research Nuclear University "MEPhI" and Project No. FSWU-2020-0035 of the Ministry of Science and Higher Education of the Russian Federation (M.H.U.).

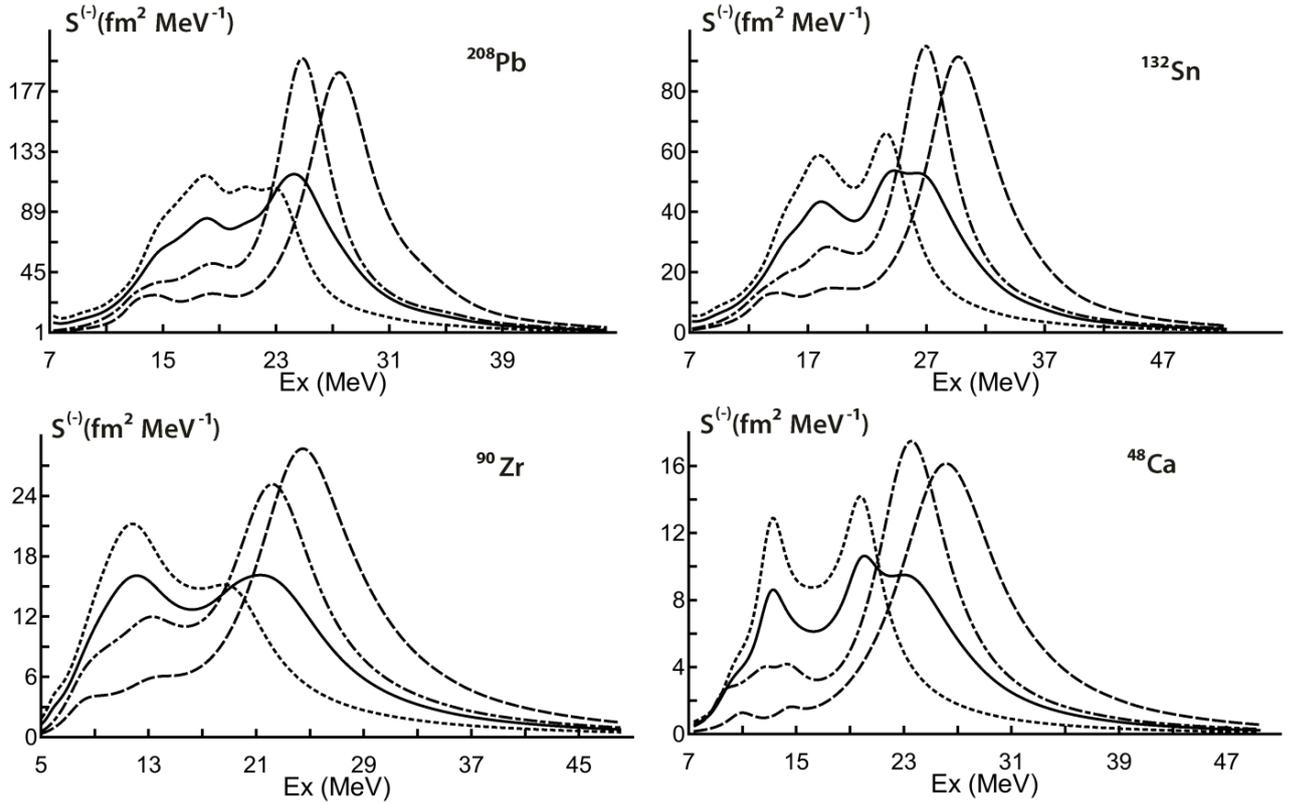

FIG.1. Evaluated within PHDOM the strength functions $S_J^{(-)}(E_x)$ of $0^-$, $1^-$ and $2^-$ J - components of IVGSDR$^{(-)}$ in parent nuclei under consideration (dashed, dashed-dotted, and dotted lines, respectively). The J-averaged strength functions $\bar{S}^{(-)}(E_x)$ (full lines) are also shown.

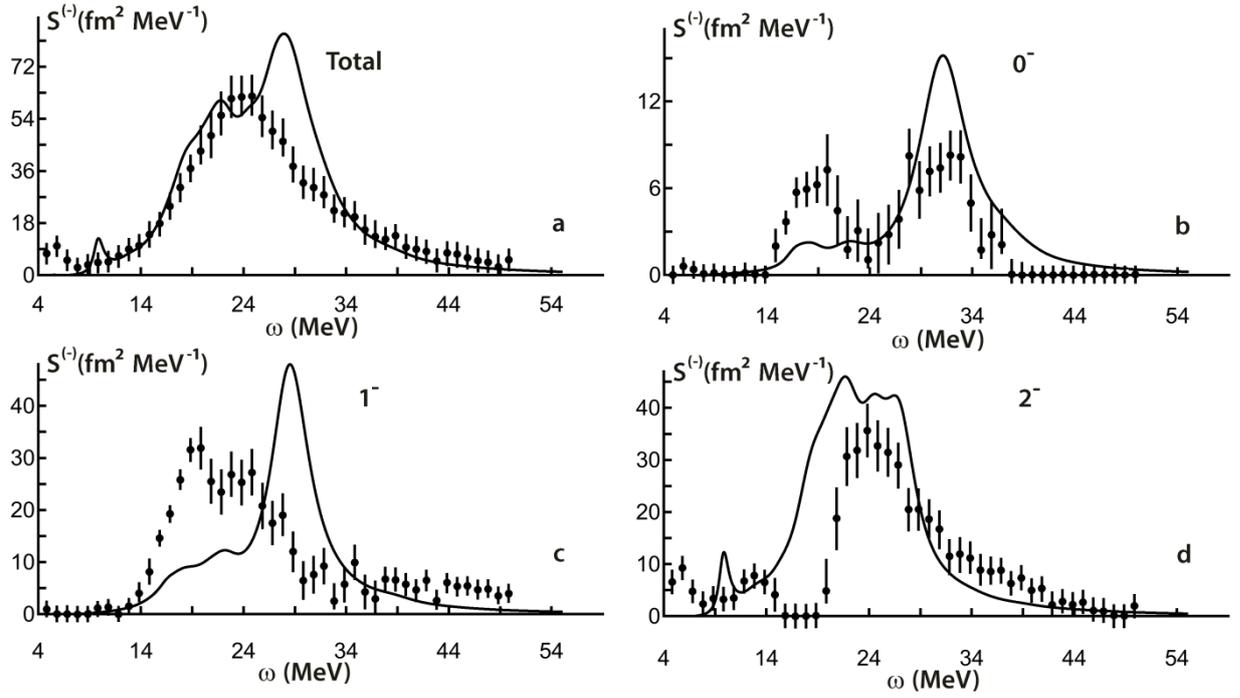

FIG. 2. Evaluated full and J-component strength functions (multiplied by $4\pi$ and $(2J+1)4\pi$, respectively) of IVGSDR$^{(-)}$ in $^{208}$Bi in a comparison with experimental data of Ref. [16].

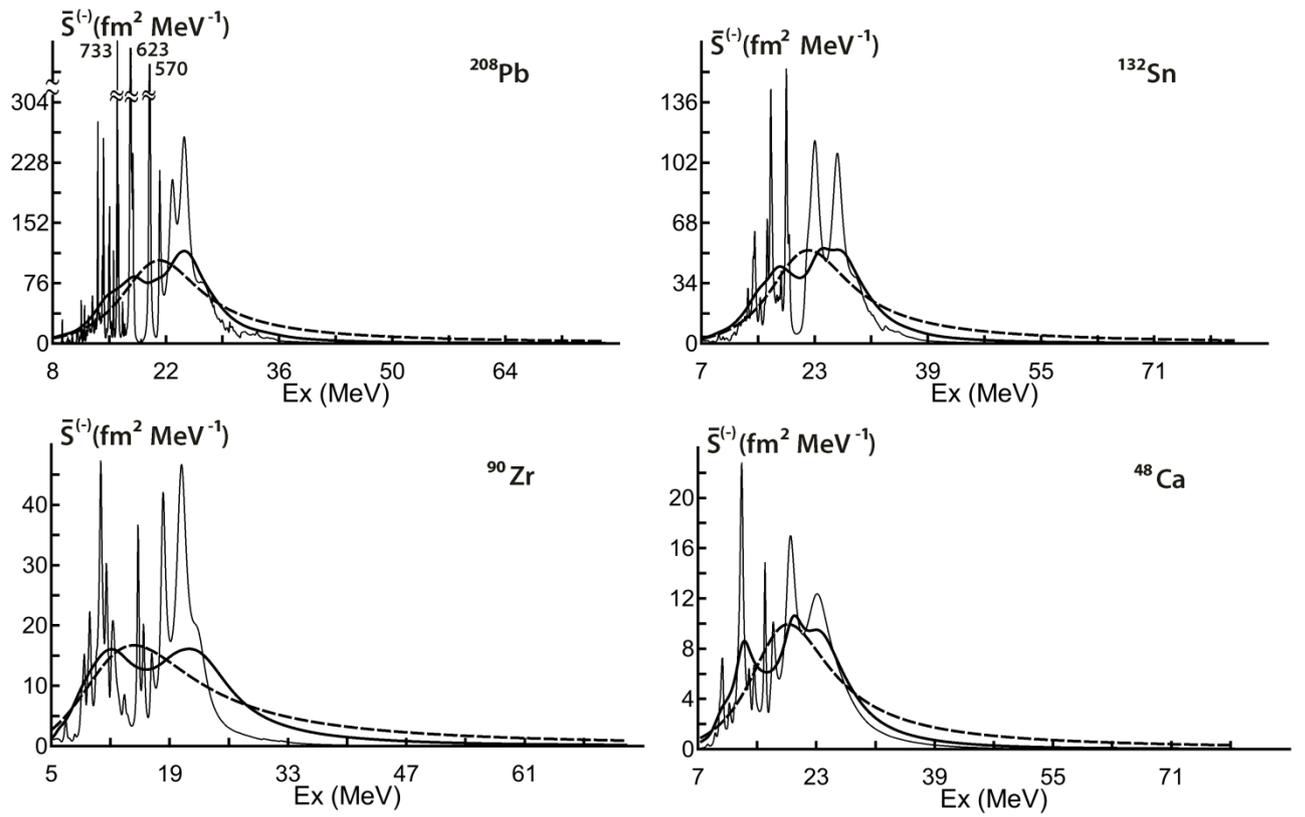

FIG.3. Evaluated J-averaged strength functions of IVGSDR$^{(-)}$ in parent nuclei under consideration (full lines) approximated by the Lorentz-type energy dependence (dashed lines). The cRPA limits of mentioned strength functions (thin lines) are also shown.

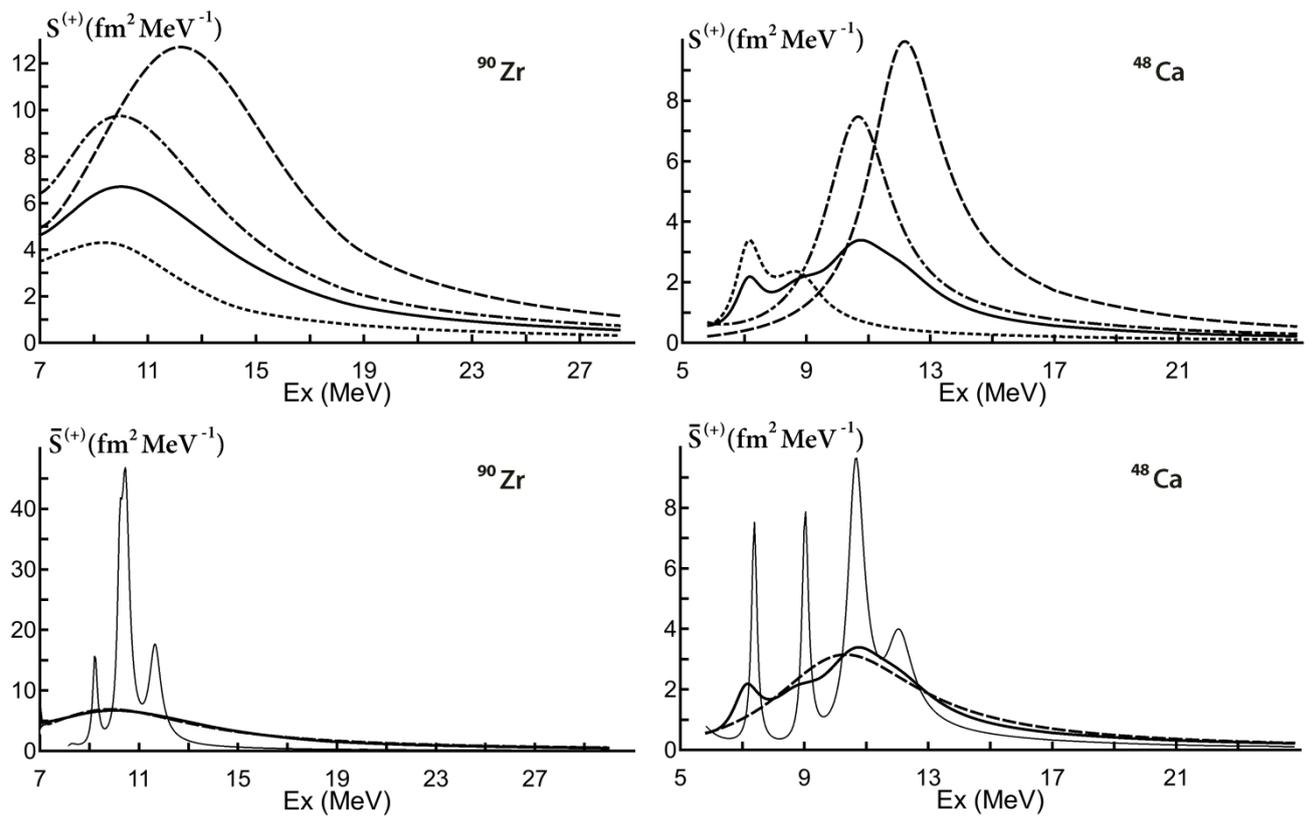

FIG.4. The same as in Figs. 1 and 3, but for IVGSDR$^{(+)}$ in $^{48}$Ca and $^{90}$Zr parent nuclei.

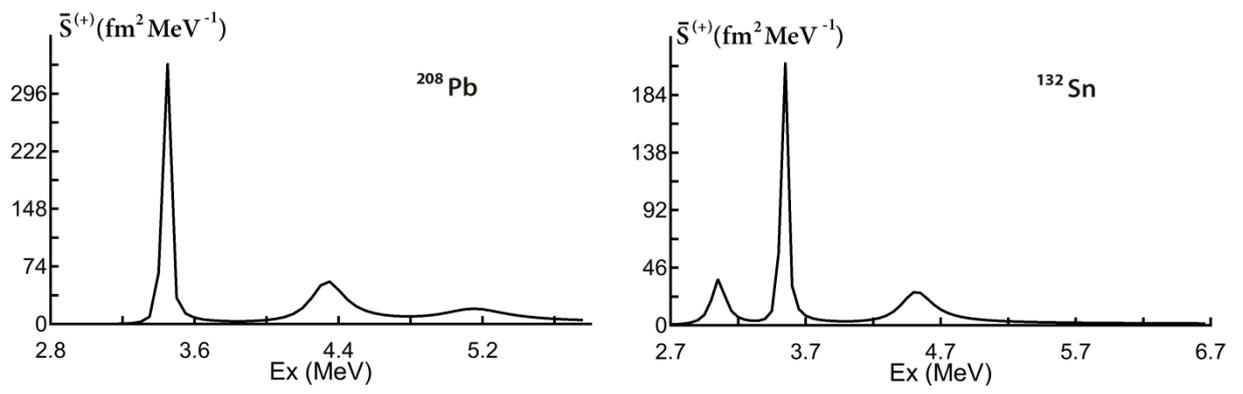

FIG.5. Evaluated J-averaged strength functions of IVGSDR$^{(+)}$ in $^{132}$Sn and $^{208}$Pb parent nuclei.

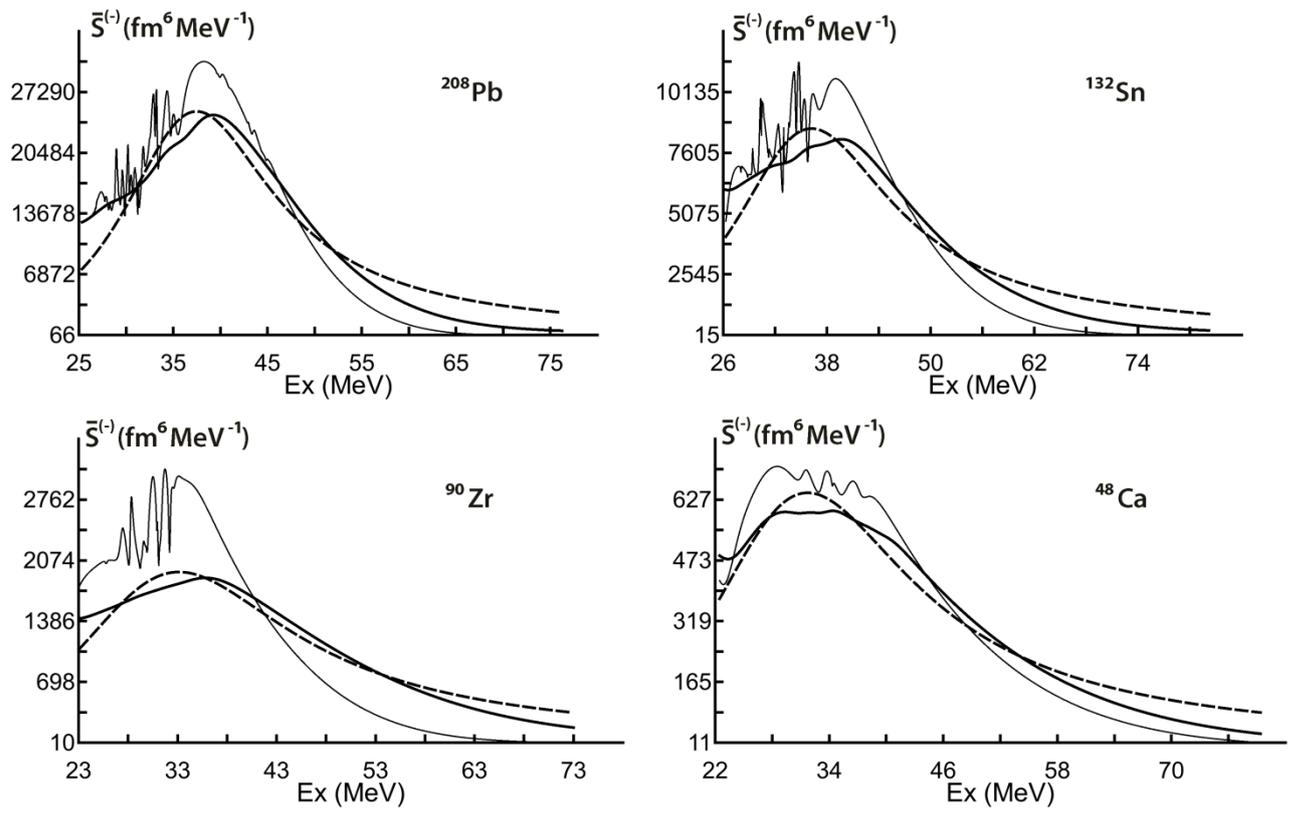

FIG.6. The same as in Fig. 3, but for IVGSDR$^{(-)}$2.

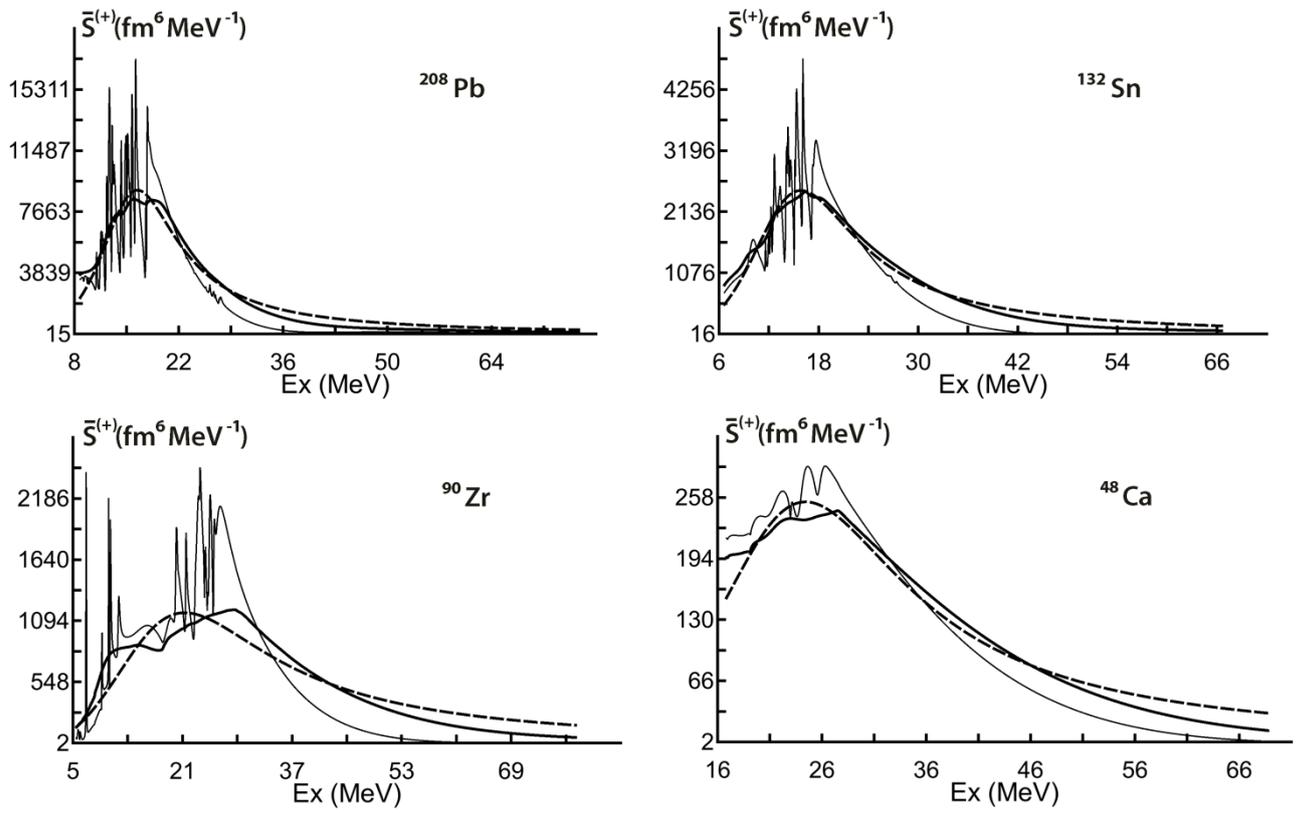

Fig.7. The same as in Fig. 4 (the second part), but for IVGSDR$^{(+)}$2 in parent nuclei under consideration.

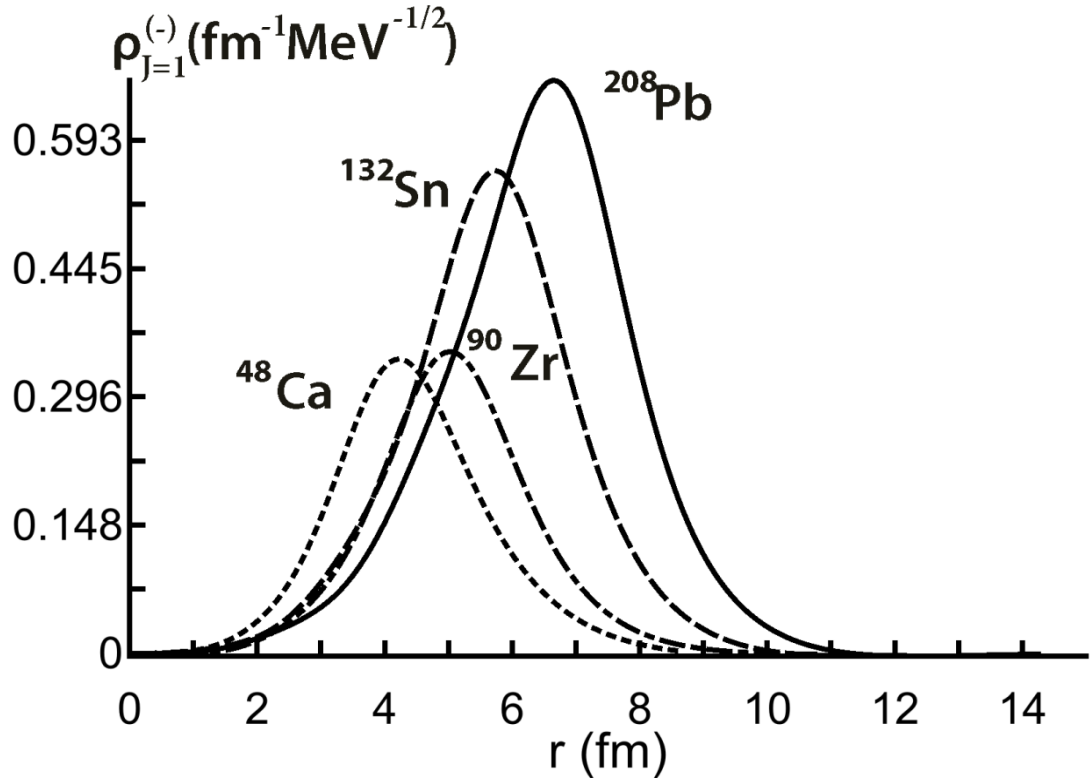

FIG.8. Taken at the excitation energy $E_x = E_{x,m_1}$, the radial element of the projected transition density, $\rho_{J=1}^{(-)}(r, E_{x,m_1})$, evaluated within PHDOM for $1^{(-)}$-IVGSDR$^{(-)}$ in parent nuclei under consideration.

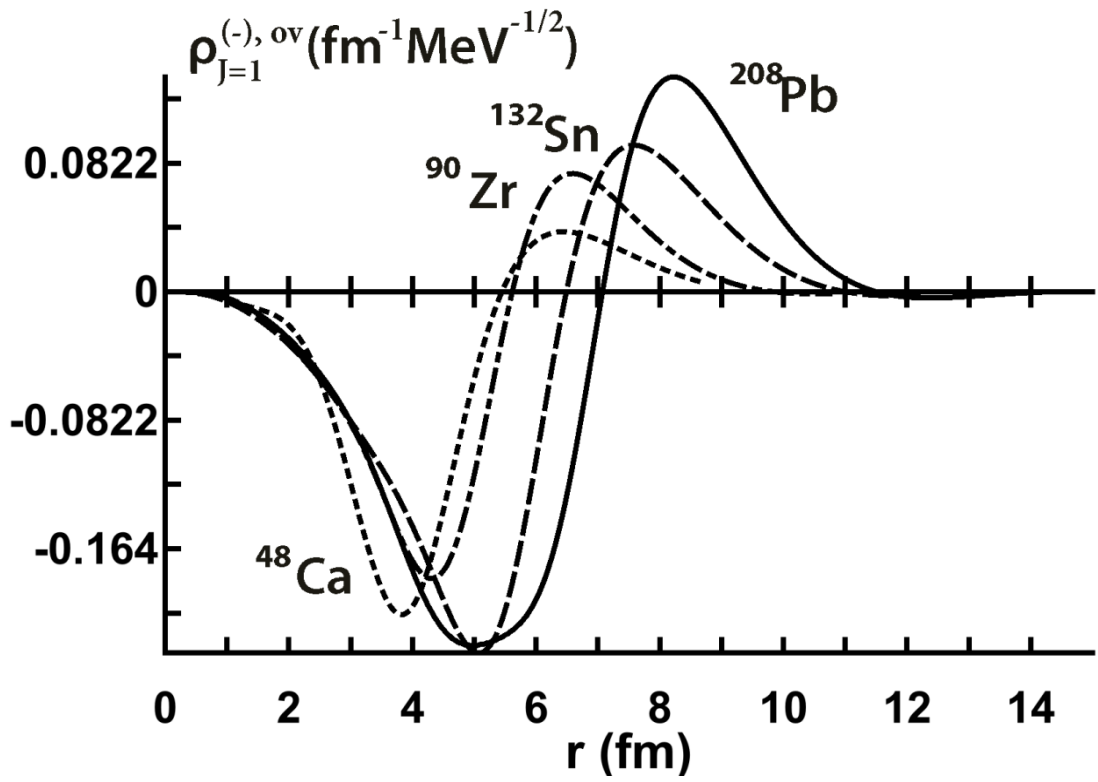

FIG.9. The same as in Fig. 8, but for $1^{(-)}$-IVGSDR$^{(-)}2$.

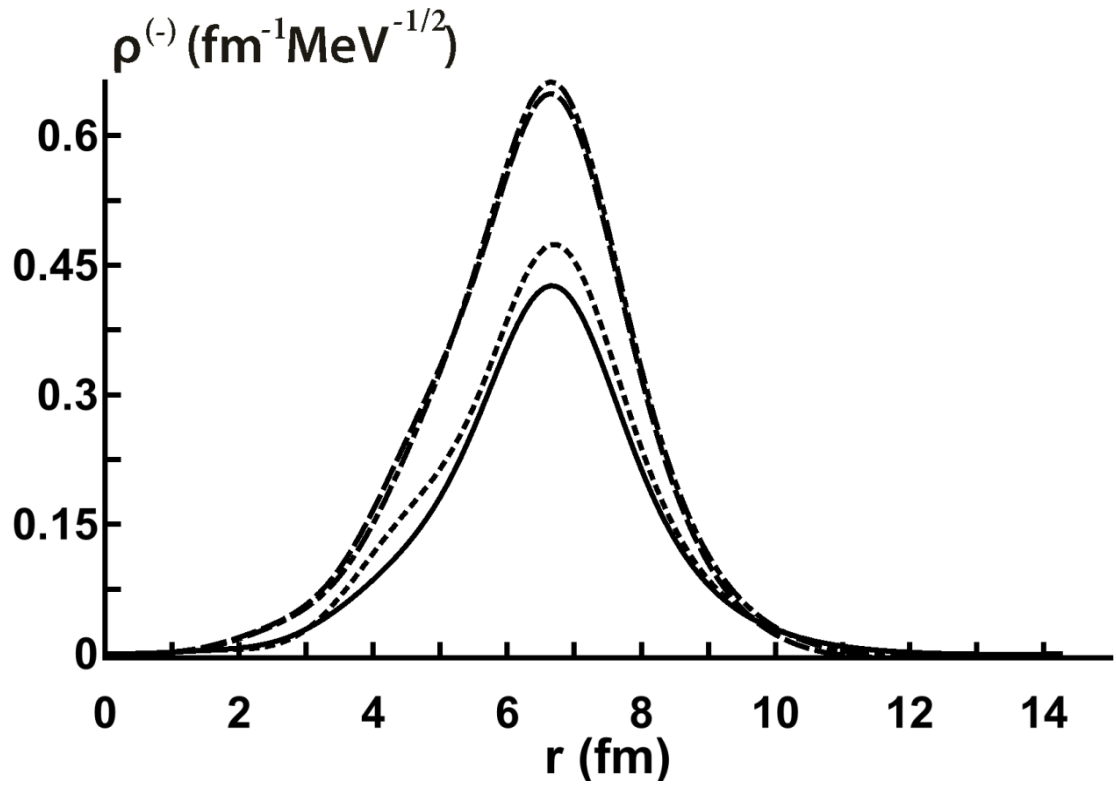

FIG.10. Radial elements of projected transition densities, $\rho_J^{(-)}(r, E_{x,m_J})$ (dashed, dashed-dotted, dotted lines for, respectively, $J = 0, 1$ and $2$) and $\bar{\rho}^{(-)}(r, E_{x,m})$ (full line), evaluated for IVGSDR$^{(-)}$ in $^{208}$Bi.

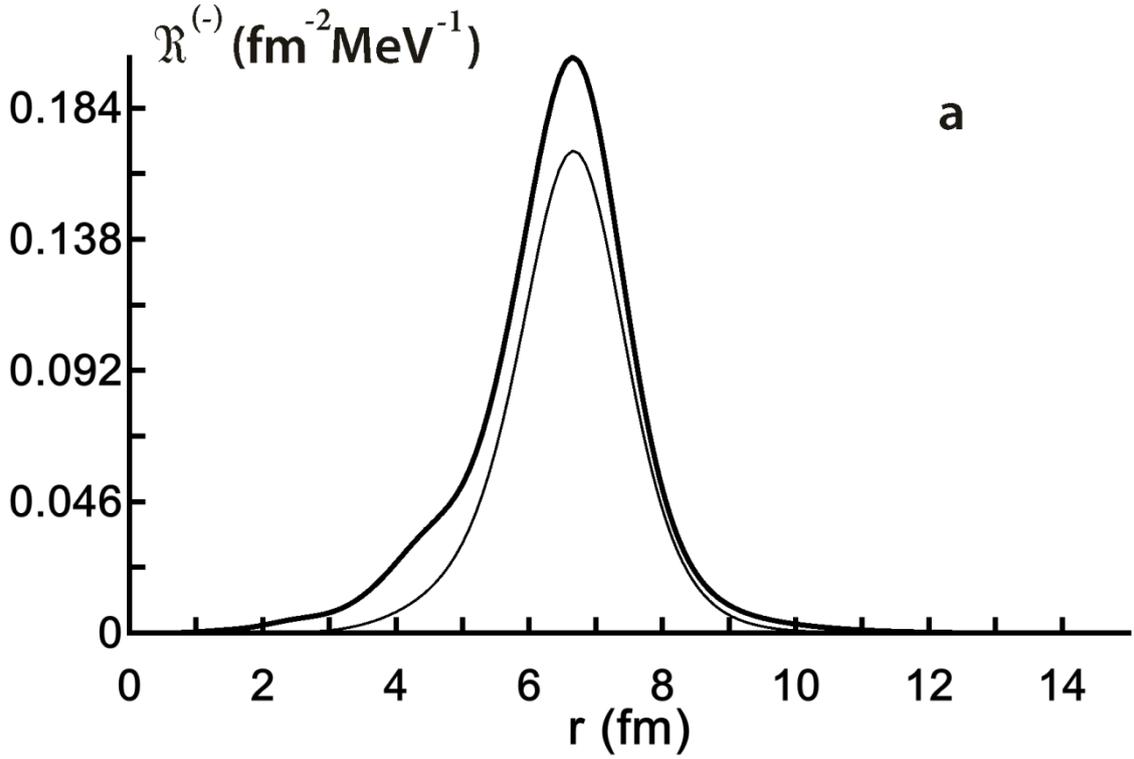

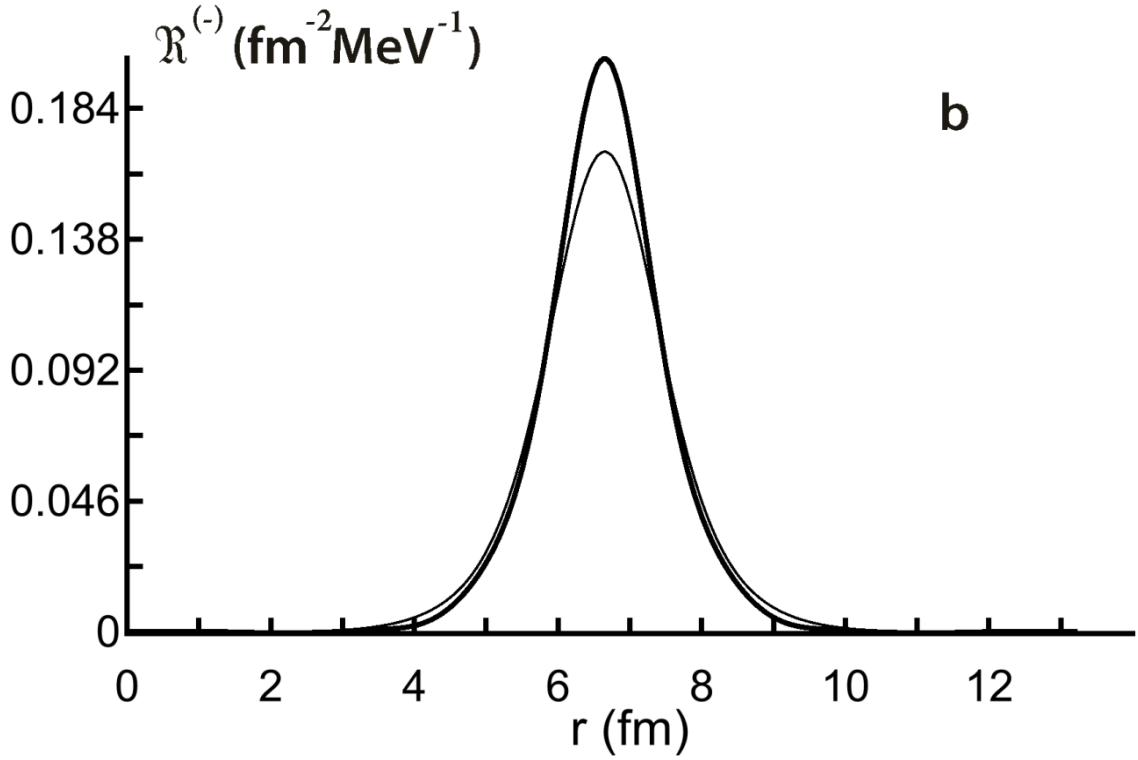

FIG.11. The radial elements of real and projected double transition densities, $\mathcal{R}^{(-)}_{J=1}(r,r',E_{x,m_1})$ and $\mathcal{R}^{(-),pr}_{J=1}(r,r',E_{x,m_1})$ (full and thin lines, respectively), taken at the energy $E_{x,m_1}$ of $1^{(-)}$-IVGSDR$^{(-)}$ in $^{208}$Bi and evaluated within PHDOM at two lines: $r = r'$ (a) and $r + r' = 2r_m$ (b).